\documentclass[aps, pre, twocolumn]{revtex4-2}
\usepackage[pdftex]{graphicx}
\usepackage{gensymb}
\usepackage{amsmath,amssymb}

\begin{document}

\title{A geometric framework for curvature-dependent collective behavior of polar active agents on curved surfaces}
\author{Tatsuo Shibata}
\affiliation{Laboratory for Physical Biology, RIKEN Center for Biosystems Dynamics Research, Kobe 650-0047, Japan}

\date{\today}

\begin{abstract}
In biological systems, active agents such as actomyosin and cells move and interact on curved surfaces, exhibiting diverse phenomena. These observations have motivated studies of how curvature shapes their collective behavior. Here, using a geometric framework, a minimal model is presented for interacting active agents on curved surfaces with Vicsek-like polar alignment. A transition between disordered and ordered states occurs on spheres as well as on oblate and prolate spheroids. As the deviation from sphericity increases, the transition point shifts to higher alignment strengths, and swarming localizes to an equatorial belt away from the poles, indicating that curvature heterogeneity influences the emergence of the polar-ordered state.
\end{abstract}

\maketitle

\section{Introduction}

Cell geometry is generally nonplanar and bounded by curved surfaces. In tissues, cell shapes often deviate from a sphere, adopting complex geometries such as squamous, cuboidal, or columnar shapes~\cite{Johnston.2011}. Even single cells migrating on flat substrates typically display a flattened geometry with a curved dorsal surface. On such curved surfaces, intracellular active agents such as actomyosin and microtubules are expected to form ordered structures driven by their motile activity, thereby contributing to both the maintenance and deformation of cell shape. Curvature-dependent emergent properties have attracted considerable attention over the past decade~\cite{Schamberger.2023,Vikran.2025,Arnold.2025,Al-Izzi.2021}. However, it remains poorly understood what kinds of collective behaviors of cytoskeletal active agents are promoted by such curved geometries. Moreover, how these curvature-induced collective behaviors contribute to the maintenance and deformation of cell shape, and thereby to tissue-scale morphogenesis, remains elusive.

\textit{In vitro} reconstitution experiments are ideally suited to studying the effect of curvature on the dynamics of cytoskeletal active agents, as they avoid the complexity of the cellular environment. For systems in which microtubules and motors, or actomyosin, are constrained to the surfaces of spheres~\cite{Keber.2014,Hsu.2022} or ellipsoids (spheroids)~\cite{Clairand.2024}, motile topological defects and a variety of structural patterns have been reported.

Tissues and organs also present curved surfaces on which cells frequently migrate. These cells can thus be regarded as active agents moving on curved surfaces. For instance, enteric neural crest cells migrate collectively along the curved surface of the developing gut~\cite{Nishiyama.2012}. Epithelial tissues on organ surfaces also often exhibit collective cell migration. For example, the intestinal epithelium displays a continuous tissue flow along the curved surface of intestinal villi toward the tip~\cite{Pérez-González.2022}. However, what kinds of collective migration modes and behaviors are induced by tissue curvature and its spatiotemporal modulation remains poorly understood. Moreover, how such curvature-induced collective behaviors contribute to tissue-scale morphogenesis remains elusive.

At the level of cell populations, \textit{in vitro} experiments also provide powerful approaches to understanding curvature-induced collective behaviors. Collective cell behaviors on curved surfaces~\cite{Xi.2017,Glentis.2022,Tang.2022} have been studied, revealing curvature-dependent collective motion. In addition, the curvature of the underlying cell substrate has been shown to influence both the direction and speed of cell migration, a phenomenon termed “curvotaxis”~\cite{Pieuchot.2018}. Furthermore, collective cell migration on the surfaces of spherical tissues has also been investigated~\cite{Brandstätter.2023}, revealing rotating collective migration accompanied by velocity waves.

These observations motivate the question of whether interacting active agents on curved surfaces exhibit distinctive ordering or pattern-forming phenomena. To address this question, hydrodynamic theories for active agents on deformable curved surfaces, such as cell surfaces and epithelial tissue monolayers~\cite{Salbreux:2017ci,Morris:2019bx,Torres-Sánchez.2022,Salbreux.2022,Bell.2022}, have been developed. Complementary to these approaches, models of interacting active agents on curved surfaces have been studied extensively~\cite{Sknepnek.2015,Ehrig.2017,Alaimo.2017,Henkes.2018,Ai.2020,Hindes.2020}, revealing a variety of swarming behaviors and dynamics of topological defects. 

For motion on curved surfaces, an intrinsic, geometry-based formulation can be employed, in which the equations of motion are posed directly on the surface using differential-geometric methods~\cite{Omori.1991,Frankel.2011}. Active, frictional, and stochastic forces can be incorporated within this framework~\cite{Fily.2016,Mackay.2025,Janzen.2025}, and interactions among active agents can be modeled by effective potentials. For coarse-grained descriptions, such as polar interactions, one must also specify how orientation vectors are defined and how they interact on a curved surface. Here, we develop a general framework to describe interacting active agents on curved surfaces and apply it to spheres as well as to oblate and prolate spheroids.

This paper is organized as follows. In Sec.~\ref{sec:singleParticle}, using a geometric framework, a model of a single active agent on curved surfaces is introduced. The distribution and dynamics of a single active agent obtained from numerical simulations are studied on spheres and spheroids. In Sec.~\ref{sec:interactingActiveParticles}, polar interacting active agents on curved surfaces are considered. Using numerical simulations,  the transition from a disordered to an ordered state on spheres and spheroids is investigated. Sec.~\ref{sec:discussion} is devoted to discussion and conclusions.

\section{Active agents on curved surfaces}
\label{sec:singleParticle}

Consider the motion of an agent on a two-dimensional curved surface $S$ endowed with a metric $g$. Let $(x,y)$ be local coordinates on $S$, so that the position of the agent is given by a point $\mathbf{r}(x,y)$ in three-dimensional space. The components of the velocity in the coordinate basis are then $(v^x,v^y) = (dx/dt,dy/dt)$. The equation of motion of the agent on $S$ is given by
\begin{equation}
    \frac{d v^\alpha}{dt} + \Gamma^\alpha_{jk} v^j v^k
    = I v^\alpha\left(v_0^2 - g_{\beta\gamma} v^\beta v^\gamma\right)
      + \sigma E^\alpha_{\;j}\,\xi^j(t),
    \label{eq:EOM_single}
\end{equation}
where summation over repeated indices is implied. 
The second term on the left-hand side is an apparent force arising from the curvature of the coordinate system,
with Christoffel symbols $\Gamma^\alpha_{jk}$ given by Eq.~(\ref{eq:defChristoffel})~\cite{Frankel.2011}.
The first term on the right-hand side is an active term that tends to maintain the speed at $v_0$ when $I>0$ is sufficiently large. A similar form of active term has been employed for active Brownian particles~\cite{Schweitzer:1998kv,Ohta:2009ga} and for migrating cells~\cite{Hiraiwa:2014fq,Hiraiwa:2019hr}. 
The second term on the right-hand side represents stochastic forcing with noise strength $\sigma$, where $E^\alpha_{\;j}$ denotes the local orthonormal basis (Eq.~(\ref{eq:localOrthoNormal})), and $\xi^j(t)$ is Gaussian white noise with zero mean and correlations $\langle \xi^i(t)\xi^j(t')\rangle=\delta^{ij}\delta(t-t')$. A heuristic derivation of Eq.~(\ref{eq:EOM_single}) is given in Appendix~\ref{appendix:EOM}.

\subsection{Active agent on sphere and spheroids}

\begin{figure}
    \includegraphics[width=.45\textwidth]{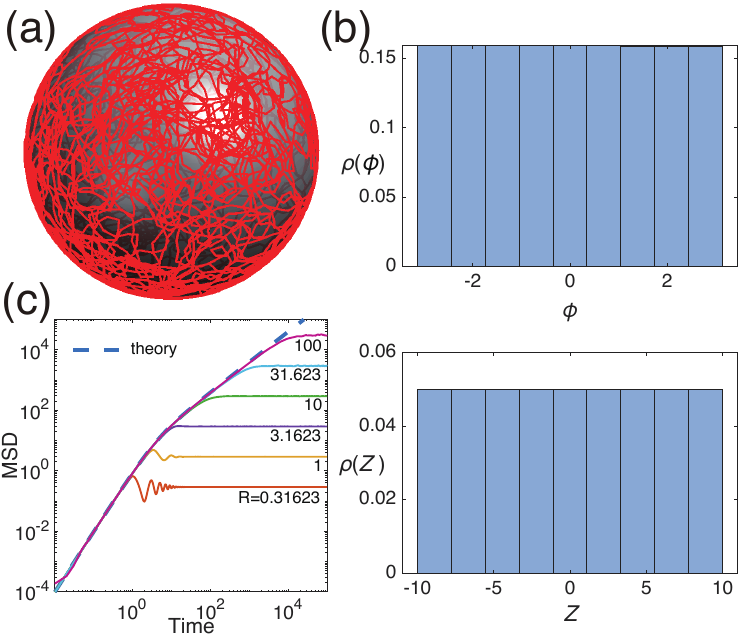}
    \caption{(a) A trajectory of an active agent on a sphere of radius $R=10$ for $t=5000$. $\sigma=1$, $v_0=1$, and $I=100$. 
    (b) The probability distributions of the azimuthal angle $\varphi$ (top)  and $Z$ (bottom). $\sigma=1$, $v_0=1$, and $I=100$.
    (c) The mean squared displacement of active agent on sphere of radius $R=\sqrt{0.1}, 1, \sqrt{10}, 10, \sqrt{1000}$, or $100$. The theory line is given by $4 \left(t + 2 \left(\exp{(-{t}/{2})} - 1\right)\right)$.}
    \label{fig:activeParticleOnSphere}
\end{figure}

First, the motion of a single active agent on a sphere of radius $R$ is considered. For a two-dimensional parametrization of the spherical surface, stereographic coordinates are used because they offer numerical simplicity and stability. In this coordinate system, each point $\mathbf{X}=(X,Y,Z)$ on the sphere embedded in three-dimensional space is assigned planar coordinates $\mathbf{x}=(x,y)$ by drawing a line from either the north or south pole through $\mathbf{X}$ and recording where it intersects the plane tangent at the south or north pole, respectively. This construction yields
$
x=\frac{R X}{R-Z},\quad y=\frac{R Y}{R-Z}
$
for projection from the north pole onto the plane tangent at the south pole, and
$
x=\frac{R X}{R+Z},\quad y=\frac{R Y}{R+Z}
$
for projection from the south pole onto the plane tangent at the north pole. In the simulations, two stereographic charts, centered at the north and south poles, are employed, and the chart is switched whenever the agent approaches the corresponding pole (see Appendix~\ref{sec:stereographic}).

Simulations were performed using Eq.~(\ref{eq:EOM_single}) with the metric $g$, Christoffel symbols $\Gamma^i_{jk}$, and local orthonormal basis $E^i_{\;j}$ of the stereographic coordinate system for a sphere of radius $R$, as given by Eqs.~(\ref{eq:metricSpheroid}), (\ref{eq:ChristoffelSpheroid}), and (\ref{eq:orthonomalSpheroid}) in Appendix~\ref{sec:stereographic}. 
For the numerical simulations, a first-order Runge--Kutta scheme for stochastic differential equations was used throughout this paper with time step $dt=0.01$~\cite{10.48550/arxiv.1210.0933}.

For sufficiently long simulations [Fig.~\ref{fig:activeParticleOnSphere}(a)], 
the probability distribution of the position approaches a uniform distribution on the sphere,
which can be identified from the probability distributions of the azimuthal angle $\varphi$ and $Z$,  
given by $\rho(\varphi)=1/2\pi$ and $\rho(Z)=1/2R$ [Fig.~\ref{fig:activeParticleOnSphere}(b)].
Next, the mean-squared displacement (MSD) is evaluated
to characterize the stochastic dynamics of an active agent on a spherical surface described by Eq.~(\ref{eq:EOM_single}).
Here, the MSD is defined as $\left\langle d\bigl(\mathbf{x}(t),\mathbf{x}(0)\bigr)^2\right\rangle$, where the distance along the shortest path between two positions on the sphere is given by $d\bigl(\mathbf{x}_1,\mathbf{x}_2\bigr)=R\theta$ with $\cos\theta={\mathbf{X}_1\cdot\mathbf{X}_2}/{R^2}$.
As shown in Fig.~\ref{fig:activeParticleOnSphere}(c), the MSD first displays a ballistic regime proportional to $t$ for $t\ll 2v_0^2/\sigma^2$, which is then followed by a random regime proportional to $t^{1/2}$ for $t\gg 2v_0^2/\sigma^2$, as in the case of an active Brownian particle.
For sufficiently long times, it shows saturation due to the finite system size.
For small $R$, this saturation occurs already for $t\ll 2v_0^2/\sigma^2$, and in such cases the MSD exhibits an oscillatory behavior due to ballistic rotational motion around the sphere.
The envelope of MSD curves for different $R$ follows the MSD of an active Brownian particle on a flat surface.
These results confirm that Eq.~(\ref{eq:EOM_single}) with appropriate $g$, $\Gamma^i_{jk}$, and $E^i_{\;j}$ describes the motion of an active agent on spheres.

\begin{figure}
    \includegraphics[width=.45\textwidth]{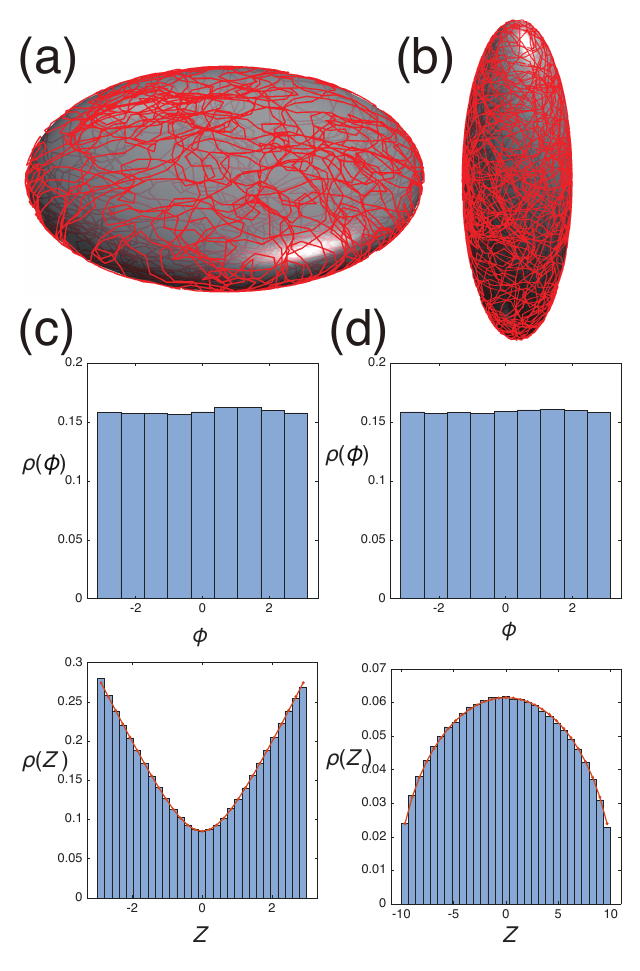}
    \caption{(a,b) Active agent on oblate and prolate spheroids of $A=10, C=3$ and $A=3, C=10$, respectively, for $t=3000$. $\sigma=1, v_0=1, I=100$.
    (c,d) The probability distributions of the azimuthal angle $\varphi$ (top)  and $Z$ (bottom) for oblate and prolate spheroids, respectively. The theoretical lines given by Eq.(\ref{eq:distribtionOfZ}) are indicated by red lines. }
    \label{fig:activeParticleOnSpheroid}
\end{figure}

Next, the motion of an active agent on spheroids is considered. A spheroid is described by
\begin{equation}
    \label{eq:spheroid}
    \frac{X^2+Y^2}{A^2}+\frac{Z^2}{C^2}=1
\end{equation}
for both the oblate ($A>C$) and prolate ($A<C$) cases. As in the case of the sphere, 
the stereographic coordinate system is used. 
The north and south pole charts are given by
$x=\frac{C X}{C-Z}$, $y=\frac{C Y}{C-Z}$ and
$x=\frac{C X}{C+Z}$, $y=\frac{C Y}{C+Z}$, respectively.
As shown in Fig.~\ref{fig:activeParticleOnSpheroid}(a) and (b), 
simulations were performed according to Eq.~(\ref{eq:EOM_single}) with 
$g$, $\Gamma^i_{jk}$, and $E^i_{\;j}$ given by Eqs.~(\ref{eq:metricSpheroid}), (\ref{eq:ChristoffelSpheroid}), and (\ref{eq:orthonomalSpheroid}), respectively, in Appendix~\ref{sec:stereographic}. 

The probability distributions of the position along a single trajectory
approach the uniform distributions on the spheroidal surface,
which can be identified from the probability distributions of the azimuthal angle $\varphi$ and $Z$, given by 
$\rho(\varphi)=1/2\pi$ and 
\begin{equation}
\rho(Z)=\frac{1}{\mathcal{N}}\sqrt{\frac{(A^2-C^2)Z^2}{C^4}+1},
\label{eq:distribtionOfZ}
\end{equation}
with the normalization constant
\[
\mathcal{N} =  A  +  \frac{C^2}{\sqrt{A^2 - C^2}} 
\operatorname{arcsinh}\!\left(\frac{\sqrt{A^2 - C^2}}{C}\right)
\quad\text{for } A>C,
\]
and
\[
\mathcal{N} =  A  +  \frac{C^2}{\sqrt{-A^2 + C^2}} 
\arcsin\!\left(\frac{\sqrt{-A^2 + C^2}}{C}\right)
\quad\text{for } A<C,
\]
[Fig.~\ref{fig:activeParticleOnSpheroid}(c) and (d)].
These results confirm that Eq.~(\ref{eq:EOM_single}) with appropriate $g$, $\Gamma^i_{jk}$, and $E^i_{\;j}$ describes the stochastic motion of active agents on spheroids.

\section{Polar interacting active agents on a curved surface}
\label{sec:interactingActiveParticles}

Now polar interacting agents are considered on a two-dimensional curved surface $S$ endowed with a metric $g$.
Let $\mathbf{x}_i=(x_i,y_i)$ be the position of agent $i$ and $\mathbf{v}_i=(v_i^x,v_i^y)$ its velocity, with 
$d\mathbf{x}_i/dt = \mathbf{v}_i$. 
For the interactions among agents, polar alignment interactions are considered. 
Then, the equation of motion of agent $i$ is given by 
\begin{equation}
\label{eq:swarm}
    \begin{aligned}
    \frac{d v^\alpha_i}{dt} + \Gamma^\alpha_{\beta\gamma} v^\beta_i v^\gamma_i  
    &= I v^\alpha_i\left(v_0^2 - g_{\beta\gamma} v_i^\beta v_i^\gamma\right) \\
    &\quad + \frac{K}{N_i}\sum_{j\in\mathcal{N}_i}\tilde{v}^\alpha_{j,i}
     + \sigma E^\alpha_{\;\beta}\,\xi^\beta_i(t),
    \end{aligned}
\end{equation}
where summation over repeated Greek indices is implied. 
The second term on the right-hand side is specific to interacting agents and represents the polar alignment interaction, where the sum runs over all agents $j$ such that the distance between $\mathbf{x}_i$ and $\mathbf{x}_j$ along the shortest path on the surface is less than the interaction distance $r$. Here, $K$ is the strength of the alignment interaction, $N_i$ is the number of agents interacting with agent $i$, and $\mathcal{N}_i$ denotes the set consisting of agent $i$ and all agents interacting with it.

Vectors tangent to the surface at different points lie in different tangent planes and therefore cannot be compared directly. To compare them intrinsically, one vector is transported along a curve on the surface to the tangent plane at the other point. This operation is called parallel transport. Specifically, to compare $\mathbf{v}_j$ at $\mathbf{x}_j$ with $\mathbf{v}_i$ at $\mathbf{x}_i$ (with $\mathbf{x}_j$ near $\mathbf{x}_i$), the vector $\mathbf{v}_j$ is parallel-transported along the (unique) length-minimizing geodesic on $S$ from $\mathbf{x}_j$ to $\mathbf{x}_i$, yielding $\tilde{\mathbf{v}}_{j,i}=(\tilde{v}^x_{j,i},\tilde{v}^y_{j,i})$, which is then compared with $\mathbf{v}_i$ at $\mathbf{x}_i$. Parallel transport along the minimizing geodesic provides a fair, coordinate-independent comparison of vectors at nearby points, introducing no distortions beyond those dictated by the curvature.

By rescaling time and spatial coordinates by the units $r/v_0$ and $r$, respectively, 
Eq.~(\ref{eq:swarm}) can be written in dimensionless form with $v_0=1$ and $r=1$. 
In this rescaled description, one obtains the dimensionless radii $R$, $A$, and $C$, 
the dimensionless coefficient $I$, 
the dimensionless alignment interaction strength $K$, 
and the dimensionless noise strength $\sigma^2$, 
which are related to the original dimensional parameters by
$R/r$, $A/r$, $C/r$, $I v_0 r$, $K r/v_0$, and $r\sigma^2 /v_0^3$, respectively.
The propulsion strength relative to random motion can be characterized by the P\'eclet number $Pe=1/\sigma^2$, and the dimensionless persistence time is $\tau_p=1/\sigma^2$.
Hereafter, the coefficient $I$ is fixed at $I=10$.

\begin{figure*}[th]
    \includegraphics[width=.8\textwidth]{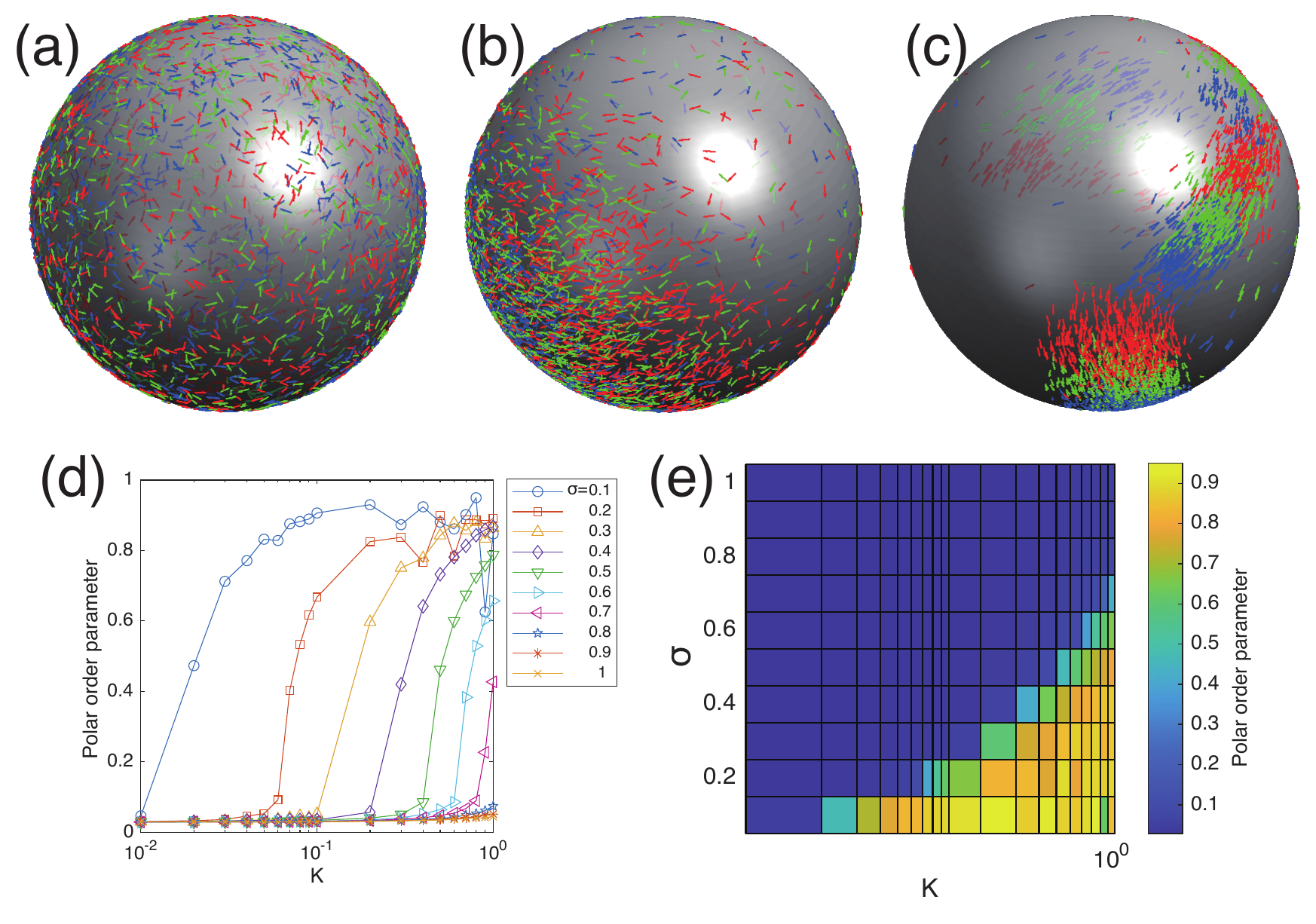}
    \caption{(a,b,c) Snapshot images of polar interacting active agents on sphere of radius $R=10$ for three different time points $t_0$ (red), $t_0+3$ (green) and $t_0+6$ (blue). Arrows indicate the velocity vectors. $N=1000$, $\sigma=0.2$ and $I=10$. 
    The interaction strength are given by (a) $K=0.05$, (b) $0.1$ and (c) $1.0$.
    (d) Polar order parameter $\lVert\mathbf{P}\rVert$ plotted against $K$ for different value of $\sigma$. 
    (e) The phase diagram in $K-\sigma$ plane showing the transition from disorder to order phases. The polar order parameter $\lVert\mathbf{P}\rVert$ is color encoded. }
    \label{fig:swarmOnSphere}
\end{figure*}

\subsection{Polar interacting agents on a sphere}

First, interacting active agents on a sphere are considered.
To calculate the distance between two agents $i$ and $j$,
the positions in stereographic coordinates $\mathbf{x}_i$ are transformed into the three-dimensional embedding coordinates $\mathbf{X}_i=(X_i,Y_i,Z_i)$,
and the distance along the geodesic (shortest path) on the sphere is computed as
$d\!\left(\mathbf{x}_i,\mathbf{x}_j\right)=R\theta_{ij}$ with $\cos\theta_{ij}={\mathbf{X}_i\cdot\mathbf{X}_j}/{R^2}$.
If $d\!\left(\mathbf{x}_i,\mathbf{x}_j\right)\leq 1$, the velocity $\mathbf{v}_j$ at $\mathbf{x}_j$ is parallel-transported to $\mathbf{x}_i$ along the geodesic, yielding $\tilde{\mathbf{v}}_{j,i}$.

For the parallel transport, Rodrigues' rotation formula is used for the velocity represented in the Euclidean coordinates, $\mathbf{V}_j$, to obtain the parallel-transported velocity
$
\tilde{\mathbf{V}}_{j,i}
=
\mathbf{V}_j\cos\theta_{ij}
+(1-\cos\theta_{ij})(\mathbf{V}_j\cdot\mathbf{n})\mathbf{n}
+(\mathbf{n}\times \mathbf{V}_j)\sin\theta_{ij},
$
with 
$\mathbf{n}={\mathbf{X}_j\times \mathbf{X}_i}/{\lvert\mathbf{X}_j\times \mathbf{X}_i\rvert}$.
The vector $\tilde{\mathbf{V}}_{j,i}$ is then transformed back into $\tilde{\mathbf{v}}_{j,i}$ in stereographic coordinates and used in Eq.~(\ref{eq:swarm}).

Simulations were performed with $N=1000$ agents on a sphere of radius $R=10$, starting from a uniform spatial distribution with random orientations.
The main parameters that control the collective behavior are the alignment strength $K$ and the noise strength $\sigma$. 
For a given $\sigma=0.2$, when $K$ is sufficiently small ($K=0.05$), the agents remain almost uniformly distributed without noticeable collective behavior [Fig.~\ref{fig:swarmOnSphere}(a) and Movie~1]. 
As $K$ is increased to $K=0.1$, the agents start to segregate into high- and low-density regions with a global polar order [Fig.~\ref{fig:swarmOnSphere}(b) and Movie~2]. 
When $K$ is sufficiently large ($K=1.0$), the agents form several clusters with polar order, moving almost in the same direction [Fig.~\ref{fig:swarmOnSphere}(c) and Movie~3]. 
Each cluster appears to move along a great circle (geodesic) of a different orientation.
For different random initial conditions, the orientation of the polar order is random, confirming that the present method does not introduce any undesirable bias. 

To quantitatively characterize the formation of polar order, a polar order parameter on the sphere is introduced as
\begin{equation}
\label{eq:polarOrderParameter}
    \mathbf{P}
    = \frac{1}{N}\sum_{i=1}^N
      \frac{\mathbf{X}_i\times\mathbf{V}_i}{\lvert\mathbf{X}_i\times\mathbf{V}_i\rvert},
\end{equation}
where $\mathbf{X}_i$ and $\mathbf{V}_i$ are the position and velocity of agent~$i$ in three-dimensional coordinates.
As shown in Fig.~\ref{fig:swarmOnSphere}(d), for a given value of $\sigma$, 
$\lVert\mathbf{P}\rVert$ increases from values close to zero to values close to one as $K$ is increased, indicating a transition from a disordered to an ordered state. 
Such a transition also occurs upon decreasing the noise strength $\sigma$ [Fig.~\ref{fig:swarmOnSphere}(e)].

\subsection{Curvature heterogeneity directs the orientation of polar active agents}

\begin{figure*}
    \includegraphics[width=.99\textwidth]{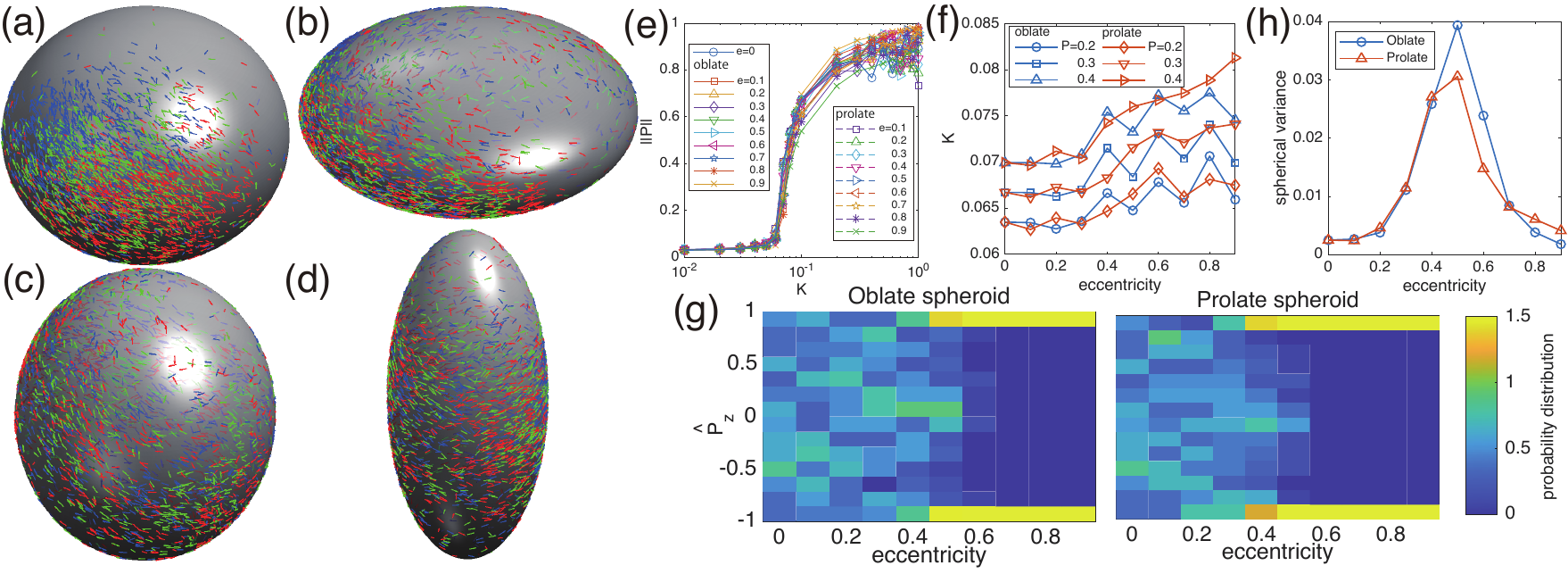}
    \caption{(a,b,c,d) Snapshot images of polar interacting active agents on oblate (a,b) and prolate (c,d) spheroids for three different time points $t_0$ (red), $t_0+3$ (green) and $t_0+6$ (blue). Arrows indicate the velocity vectors. $N=1000$, $K=0.2$, $\sigma=0.2$, $R=10$, and $I=10$. 
    The eccentricity $e$ is given by (a,c) $e=0.5$ and (b,d) $e=0.9$. 
    (e) Polar order parameter $\lVert\mathbf{P}\rVert$ plotted against $K$ for oblate and prolate spheroids with different value of eccentricity $e$. $\sigma=0.2$, $N=1000$. 
    (f) The $K$-values that give $\lVert\mathbf{P}\rVert=0.2, 0.3, 0.4$ are plotted against eccentricity $e$. Interpolation was used to obtain the $K$-values. 
    (g) The distribution of $\hat{P}_Z$ for different value of eccentricity $e$ for oblate and prolate spheroids. $K=0.1$, $\sigma=0.1$, $N=1000$, $R=10$ and $I=10$. The distributions are obtained from 100 samples. 
    (h) The spherical variance of the temporal fluctuation in the unit polar order vector $\hat{\mathbf{P}}$ averaged over 100 samples plotted against eccentricity $e$ for oblate and prolate spheroids. $K=0.1$, $\sigma=0.1$, $N=1000$, $R=10$ and $I=10$.}
    \label{fig:swarmOnSpheroid}
\end{figure*}

A sphere is homogeneous and isotropic in curvature, with Gaussian curvature $1/R^2$.
To examine how curvature inhomogeneity and anisotropy affect the collective behavior of polar interacting active agents, oblate and prolate spheroids, given by Eq.~(\ref{eq:spheroid}), are considered as simple examples. 
The deviation from sphericity is specified by the eccentricity $e$, defined as
$e=\sqrt{1-{C^2}/{A^2}}$ for oblate spheroids ($A>C$) 
and
$e=\sqrt{1-{A^2}/{C^2}}$ for prolate spheroids ($A<C$).
To keep the agent density constant, the semi-axes $A$ and $C$ are adjusted so that the surface area remains $S=4\pi R^2$.
The explicit forms of $A$ and $C$ are given in Appendix~\ref{app:spheroid}.

On spheroids, simple analytical expressions for the geodesic distance between two points and for the parallel transport of a vector along a geodesic are not available in closed form. Approximate expressions are used instead~\cite{Brewin.20096c}. 
The derivation is given in Appendix \ref{app:approximation}.
The distance between two points $\mathbf{x}_i$ and $\mathbf{x}_j$ is approximately given by 
\begin{equation}
    d^2(\mathbf{x}_i,\mathbf{x}_j)
    = g_{\alpha\beta}(\mathbf{m})\,\Delta^\alpha\Delta^\beta
      + O(\lVert\Delta\rVert^4),
      \label{eq:approximateDistance}
\end{equation}
where $\Delta^\alpha = x_j^\alpha - x_i^\alpha$ and 
the metric $g_{\alpha\beta}(\mathbf{m})$ is evaluated at the midpoint 
$\mathbf{m}=(\mathbf{x}_i+\mathbf{x}_j)/2$. 
The approximate expression for the parallel transport of a vector $\mathbf{v}_j$ at $\mathbf{x}_j$ to $\mathbf{x}_i$ along the geodesic between them is given by
\begin{equation}
        \begin{aligned}
\tilde{v}_{j,i}^\alpha
=
\left(
\delta_\beta^\alpha
+\Gamma^\alpha_{\beta k}(\mathbf{m})\Delta^k
+\frac{1}{2}\Gamma^\alpha_{\gamma k}(\mathbf{m})\Gamma^\gamma_{\beta \ell}(\mathbf{m})\Delta^k\Delta^\ell
\right)
v_j^\beta
\\
+O(\lVert\Delta\rVert^3),
        \end{aligned}
        \label{eq:parallelTransportApproximation}
\end{equation}
where the Christoffel symbols $\Gamma^\alpha_{\beta k}(\mathbf{m})$ are evaluated at the midpoint $\mathbf{m}$.

Simulations were performed with $N=1000$ agents on spheroids with $R=10$, starting from a uniform spatial distribution with random orientations. The collective behavior depends on the shape of the surface for the same values of $K$ and $\sigma$ [Fig.~\ref{fig:swarmOnSpheroid}(a)–(d), Movies~4–7, $K=0.2$ and $\sigma=0.2$]. 
Swarms on oblate spheroids [Fig.~\ref{fig:swarmOnSpheroid}(a) and (b), Movies~4 and 5] appear more concentrated than those on prolate spheroids [Fig.~\ref{fig:swarmOnSpheroid}(c) and (d), Movies~6 and 7]. 
For larger values of the eccentricity $e$ [Fig.~\ref{fig:swarmOnSpheroid}(b) and (d)], the swarms tend to form in the equatorial region. 

First, the effect of surface shape on the disorder–order transition was examined.  
In Fig.~\ref{fig:swarmOnSpheroid}(e), the polar order parameter $\lVert\mathbf{P}\rVert$ given by Eq.~(\ref{eq:polarOrderParameter}) is plotted as a function of $K$ for different surface shapes with different values of $e$ for both oblate and prolate cases. 
As $K$ is increased, $\lVert\mathbf{P}\rVert$ increases from values close to zero to values close to one, indicating a transition from a disordered to an ordered state. 
In the ordered phase for sufficiently large $K$,  
$\lVert\mathbf{P}\rVert$ shows only a slight dependence on $e$.  
Moreover, the transition point from disorder to order shifts to larger values of $K$ as $e$ is increased [Fig.~\ref{fig:swarmOnSpheroid}(f)], 
indicating that a stronger alignment interaction is required on surfaces with curvature heterogeneity to induce a polar-ordered state.

Next, the influence of the surface shape on the direction and spatial region of swarm formation was investigated for $K=0.1$ and $\sigma=0.2$. 
The polar order parameter $\mathbf{P}$ is almost perpendicular to the plane in which the swarm is formed. 
If this plane is oriented in random directions, the $Z$ component of 
the unit polar order parameter $\hat{\mathbf{P}}=\mathbf{P}/\lVert \mathbf{P} \rVert$,
namely $\hat{P}_Z$, is expected to be uniformly distributed, whereas any bias in its orientation should manifest as deviations from a uniform distribution. 
For a single simulation, $\hat{P}_Z$ reaches a steady-state value with fluctuations after an initial transient.
Therefore, multiple simulations ($n=100$) were performed to obtain the distribution of $\hat{P}_Z$.
For the sphere with $e=0$, $\hat{P}_Z$ is distributed almost uniformly, as shown in Fig.~\ref{fig:activeParticleOnSpheroid}(f).
In contrast, for eccentricities $e$ larger than about $0.5$ in both oblate and prolate spheroids, $\hat{P}_Z$ is biased toward values close to $1$ and $-1$, indicating that the swarms form preferentially in the equatorial region [Fig.~\ref{fig:swarmOnSpheroid}(f)]. 
In the intermediate range around $e=0.4$, a small peak around $\hat{P}_Z\approx 0$ appears, corresponding to swarm formations moving through the two poles. 
These results indicate that curvature inhomogeneity and anisotropy affect the collective behavior of polar interacting active agents.

The fluctuation in the orientation of $\hat{\mathbf{P}}$ 
shows a dependence on the eccentricity $e$.
To quantify this dependence, the temporal variance of $\hat{\mathbf{P}}$ was measured. 
As the orientational fluctuation increases, 
the magnitude of the temporal average $\langle\hat{\mathbf{P}}(t)\rangle$ decreases.
The spherical variance $V_{\textrm{sph}}$ is then defined as  
$
V_{\textrm{sph}} = 1-\lVert\langle\hat{\mathbf{P}}(t)\rangle\rVert.
$
As shown in Fig.~\ref{fig:swarmOnSpheroid}(g), 
$V_{\textrm{sph}}$ increases as the curvature inhomogeneity increases, 
indicating that the stability of the polar-order direction decreases.
However, upon a further increase in curvature inhomogeneity, 
$V_{\textrm{sph}}$ starts to decrease, indicating that the direction of polar order $\hat{\mathbf{P}}$ becomes stabilized along the polar axis.

\section{Conclusion}
\label{sec:discussion}

In this paper, a model of interacting active agents on curved surfaces is proposed using a geometric framework. 
Trajectories of single active agents on a sphere and on spheroids approach uniform distributions along the curved surface. 
A Vicsek-like polar alignment interaction is introduced by using parallel transport of velocity vectors between nearby positions. 
As the alignment strength is increased or the noise strength is decreased, a transition from a disordered to an ordered state occurs on both spheres and spheroids. 
The transition point depends on the eccentricity, suggesting that a stronger alignment effect is needed to induce a polar-ordered state on surfaces with curvature heterogeneity.

When the curvature is uniform or nearly uniform, swarming behavior occurs in almost random directions. 
However, as the deviation from a sphere is increased in both oblate and prolate spheroids, polar-ordered swarming occurs along the equator, away from the two poles, indicating that curvature heterogeneity affects the formation of polar order. 
The formation of collective behavior along the equatorial region with defects localized around the poles has been reported previously on spheroids in a model of polar interacting active agents~\cite{Ehrig.2017} and in experiments of microtubules with nematic interaction~\cite{Clairand.2024}. 
For intermediate curvature heterogeneity, the orientation of swarming behavior exhibits larger temporal fluctuations.

Curvature can affect alignment interactions~\cite{Mackay.2025}.
Even if two agents move in parallel at nearby positions, curvature eventually changes their directions of motion, thereby reducing the effect of alignment. 
In the absence of noise, individual agents move along geodesics, and swarms also tend to move approximately along geodesic paths. 
In regions where geodesic trajectories change significantly when the initial position is shifted slightly in the direction transverse to the trajectory, it is difficult to maintain coherent swarming motion. 
On spheroids, such curvature-induced decorrelation of alignment is minimized along the equatorial region, which tends to orient the swarming motion along the equator.

The present framework provides a minimal, geometry-based description of polar active swarms on curved surfaces. 
In this study, interacting active agents were considered on spheres and spheroids. However, the method is applicable to arbitrary curved surfaces once the metric is specified.
Thus, it may be used to interpret curvature-dependent collective behaviors
observed in biological systems, such as epithelial tissues on curved organs, cell populations migrating along intestinal villi, or cytoskeletal structures on curved cell surfaces.
Extending this approach to deformable surfaces would allow more realistic modeling of cellular and tissue dynamics on complex geometries.

\begin{acknowledgments}
I thank Biplab Bhattacherjee for critical reading of this manuscript and the members of Laboratory for Physical Biology, RIKEN BDR for discussions. 
This work was supported by JSPS KAKENHI (22H05170), 
and core funding at RIKEN Center for Biosystems Dynamics Research.
\end{acknowledgments}

\appendix

\section{Derivation of the equation of motion for an active agent on  curved surfaces}
\label{appendix:EOM}

In this section, the coordinates $x^1$ and $x^2$ are used instead of $x$ and $y$ in the main text.
Suppose that $\mathbf{r}(x^1,x^2)$ is a curved surface $S\subset\mathbb{R}^3$ with coordinates $(x^1, x^2)$ on the surface. 
Then, the vectors $\frac{\partial \mathbf{r}}{\partial x^i}$ are tangent to the surface $S$. 
The unit normal vector to the surface, $\mathbf{n}(x^1,x^2)$, is given by 
\begin{equation}
\mathbf{n}=
\frac{
\frac{\partial \mathbf{r}}{\partial x^1}\times\frac{\partial \mathbf{r}}{\partial x^2}
}{
\left\lVert\frac{\partial \mathbf{r}}{\partial x^1}\times\frac{\partial \mathbf{r}}{\partial x^2}\right\rVert
}.
\end{equation}
The metric on $S$ is given by 
\begin{equation}
g_{ij}
=
\frac{\partial \mathbf{r}}{\partial x^i}
\cdot
\frac{\partial \mathbf{r}}{\partial x^j}.
\label{eq:metric}
\end{equation}

The equation of motion of an agent on the surface $S$ may be written as 
\begin{equation}
    \frac{d^2\mathbf{r}(x^1(t),x^2(t))}{dt^2}
    = \mathbf{F}(\mathbf{r},\dot{\mathbf{r}}),
    \label{eq:3D_EOM}
\end{equation}
where $\mathbf{F}(\mathbf{r},\dot{\mathbf{r}})$ is the force acting on the agent on $S$. 
The left-hand side can be expanded as
\begin{equation}
    \frac{d^2\mathbf{r}}{dt^2}=
    \frac{\partial \mathbf{r}}{\partial x^i}
    \frac{d^2x^i}{dt^2}
    +
    \frac{\partial^2 \mathbf{r}}{\partial x^i \partial x^j}
    \frac{dx^i}{dt}
    \frac{dx^j}{dt}.
\end{equation}
The first term on the right-hand side is tangential to the surface. 
The second term can be decomposed into tangential and normal components as
\begin{equation}
\frac{\partial^2 \mathbf{r}}{\partial x^i \partial x^j}=
\frac{\partial \mathbf{r}}{\partial x^k}\Gamma^k_{ij}
+\mathbf{n}\,h_{ij},
\label{eq:decomposition}
\end{equation}
where $\Gamma^k_{ij}$ is the Christoffel symbol, which is related to the metric $g_{ij}$. 
From Eqs.~(\ref{eq:metric}) and (\ref{eq:decomposition}), one obtains
\begin{equation}
   \frac{\partial g_{ij}}{\partial x^k} 
   = g_{\ell j} \Gamma^\ell_{ik} + g_{i \ell} \Gamma^\ell_{jk}.
   \label{eq:metricDerivative}
\end{equation}
Using $g_{ij}=g_{ji}$, $\Gamma^\ell_{ik}=\Gamma^\ell_{ki}$, 
and the inverse metric $g^{ij}$ of $g_{ij}$, the Christoffel symbols $\Gamma^i_{jk}$ are obtained from the metric $g$ as
\begin{equation}
\Gamma^i_{jk}
=
\frac{1}{2}g^{i\ell}
\left(
\frac{\partial g_{\ell k}}{\partial x^j}
+\frac{\partial g_{\ell j}}{\partial x^k} 
- \frac{\partial g_{jk}}{\partial x^\ell}
\right).
\label{eq:defChristoffel}
\end{equation}

The force $\mathbf{F}(\mathbf{r},\dot{\mathbf{r}})$ on the right-hand side of Eq.~(\ref{eq:3D_EOM}) can be decomposed into a component driving the motion along the surface, $\frac{\partial \mathbf{r}}{\partial x^i} f^i$, and a component that constrains the agent to the surface, such as an adhesion force, $\lambda \mathbf{n}$. Thus,
\begin{equation}
\mathbf{F}(\mathbf{r},\dot{\mathbf{r}})
=
\frac{\partial \mathbf{r}}{\partial x^i} f^i
+ \lambda \mathbf{n}.
\end{equation}

The tangential components in Eq.~(\ref{eq:3D_EOM}) are of interest here. 
From the above decompositions on both sides of Eq.~(\ref{eq:3D_EOM}),
the equation of motion along the surface is given by
\begin{equation}   
\frac{d^2x^k(t)}{dt^2}
+
\Gamma^k_{ij}
\frac{dx^i}{dt}
\frac{dx^j}{dt}
=
f^k.
\end{equation}

For the force tangential to the surface $S$ that drives the agent along it, 
an active force maintaining self-propulsion is considered as 
\begin{equation}
    \frac{\partial \mathbf{r}}{\partial x^\ell} f^\ell
    =
    I\left(v_0^2-\lvert\dot{\mathbf{r}}\rvert^2\right)\dot{\mathbf{r}}.
\end{equation}
When $I>0$, active motion is maintained. 
A similar formulation has been employed previously~\cite{Schweitzer:1998kv,Ohta:2009ga,Hiraiwa:2014fq,Hiraiwa:2019hr}. 
Using $\dot{\mathbf{r}}=\dot{x}^i\,\frac{\partial \mathbf{r}}{\partial x^i}$, we obtain
\begin{equation}
    f^k
    =
    I \left(v_0^2-g_{ij}\dot{x}^i\dot{x}^j\right)\dot{x}^k.
\end{equation}

The noise contribution to the force $\mathbf{F}(\mathbf{r},\dot{\mathbf{r}})$ in Eq.~(\ref{eq:3D_EOM}) is also considered. 
First, a noise term in three-dimensional space, $\mathbf{\Xi}=(\xi_x,\xi_y,\xi_z)$, is introduced, where the components are Gaussian white noises with zero mean and unit variance,
$
\bigl\langle \xi_i(t)\xi_j(t')\bigr\rangle
=\delta_{ij}\,\delta(t-t').
$
The contribution to the motion along the surface is then given by 
$
f^k=\sigma\, g^{kn}\,\frac{\partial \mathbf{r}}{\partial x^n}\cdot\mathbf{\Xi},
$
with noise strength $\sigma$ and inverse metric $g^{ij}$.

Next, stochastic variables $(\xi^1, \xi^2)$ are introduced as
\begin{equation}
\begin{cases}
    \xi^1
=
\displaystyle
\frac{1}{\sqrt{g_{11}}}
\frac{\partial \mathbf{r}}{\partial x^1}
\cdot
\mathbf{\Xi},
\\[0.8em]
\xi^2
=
\displaystyle
\frac{
-g_{12}\,\frac{\partial \mathbf{r}}{\partial x^1}
+g_{11}\,\frac{\partial \mathbf{r}}{\partial x^2}
}{\sqrt{g_{11}}\sqrt{g_{11}g_{22}-g_{12}^2}}
\cdot
\mathbf{\Xi},
\end{cases}
\end{equation}
which are also Gaussian white noises with zero mean and unit variance,
$
\langle \xi^i(t)\rangle=0,
$
$
\langle\xi^i(t)\xi^j(t')\rangle=\delta^{ij}\,\delta(t-t').
$
Then, the noise contribution to the motion along the surface can be written as 
\begin{equation}
    f^k
=
\sigma\,E^k_j\,\xi^j
\end{equation}
with the local orthonormal basis 
\begin{equation}
    \label{eq:localOrthoNormal}
    \begin{cases}
(E_1^1,E_1^2) =\displaystyle\frac{1}{\sqrt{g_{11}}}(1,0),\\[0.6em]
(E_2^1,E_2^2)=
\displaystyle
\frac{1}{\sqrt{g_{11}}\sqrt{g_{11}g_{22}-g_{12}^2}}
(-g_{12},g_{11}).
    \end{cases}
\end{equation}
With $v^k=dx^k/dt$, the equation of motion of an active agent on a curved surface is finally obtained in the form of Eq.~(\ref{eq:EOM_single}).

\section{Stereographic coordinate system for sphere and spheroid}
\label{sec:stereographic}

Here, the metric $g$, the Christoffel symbols $\Gamma^i_{jk}$, and the local orthonormal basis $E^i_{\;j}$ of the spheroid given by Eq.~(\ref{eq:spheroid}) in the stereographic coordinate system are presented. 
Those for a sphere are obtained by setting $A=C=R$ in the following expressions. 

The metric $g$ of the spheroid is given by
\begin{equation}
    \label{eq:metricSpheroid}
\begin{cases}
g_{xx}=\dfrac{4A^4}{r_2^2}-\dfrac{16 A^4 \Delta }{r_2^4}x^2,\\[0.4em]
g_{yy}=\dfrac{4A^4}{r_2^2}-\dfrac{16 A^4 \Delta }{r_2^4}y^2,\\[0.4em]
g_{xy}=g_{yx}=-\dfrac{16 A^4 \Delta }{r_2^4}xy,
\end{cases}  
\end{equation}
with 
$
r_2=x^2+y^2+A^2,
$
$
\Delta = A^2-C^2.
$

The Christoffel symbols are given by
\begin{equation}
    \label{eq:ChristoffelSpheroid}
    \begin{cases}
\Gamma^x_{xx}=-\dfrac{2\left(r_2^2+2\Delta(A^2-3x^2-3y^2)\right)}{r_2\beta}\,x,\\[0.4em]
\Gamma^x_{xy}=\Gamma^x_{yx}=-\dfrac{2y}{r_2},\\[0.4em]
\Gamma^x_{yy}=\dfrac{2(r_2-2\Delta)}{\beta}\,x,\\[0.4em]
\Gamma^y_{xx}=\dfrac{2(r_2-2\Delta)}{\beta}\,y,\\[0.4em]
\Gamma^y_{xy}=\Gamma^y_{yx}=-\dfrac{2x}{r_2},\\[0.4em]
\Gamma^y_{yy}=-\dfrac{2\left(r_2^2+2\Delta(A^2-3x^2-3y^2)\right)}{r_2\beta}\,y,
    \end{cases}
\end{equation}
with $\beta = r_2^2-4\Delta (x^2+y^2)$.

The local orthonormal basis vectors are given by
\begin{equation}
    \label{eq:orthonomalSpheroid}
E_1 = \left(\frac{r_2^2}{2A^2\sqrt{\alpha}},\,0\right),\qquad
E_2 = \left(\frac{2 r_2 \Delta }{A^2\sqrt{\alpha\beta}}xy,\,
\frac{r_2\sqrt{\alpha}}{2A^2 \sqrt{\beta}}\right),
\end{equation}
with $\alpha = r_2^2-4\Delta x^2$.

North and south pole charts are considered. In the numerical simulations, switching between charts is performed when an agent is at the equator, $x^2 + y^2 = A^2$.  
To switch the coordinates from the south pole chart $(x_S, y_S,  v_{x_S}, v_{y_S})$ to the north pole chart 
$(x_N, y_N, v_{x_N}, v_{y_N})$, the following formulae are used:
\begin{equation}
\begin{cases}
    x_N = \dfrac{A^2 x_S}{x_S^2 + y_S^2},\qquad
    y_N = \dfrac{A^2 y_S}{x_S^2 + y_S^2},\\[0.6em]
    v_{x_N} = 
    -A^2 \dfrac{x_S^2 - y_S^2}{(x_S^2 + y_S^2)^2}\, v_{x_S}
    - 2 A^2 \dfrac{x_S y_S}{(x_S^2 + y_S^2)^2}\, v_{y_S},\\[0.6em]
    v_{y_N} = 
    - 2 A^2 \dfrac{x_S y_S}{(x_S^2 + y_S^2)^2}\, v_{x_S}
    + A^2 \dfrac{x_S^2 - y_S^2}{(x_S^2 + y_S^2)^2}\, v_{y_S}.
\end{cases}
\end{equation}
The same formulae hold for switching from the north to the south pole charts, with the subscripts $N$ and $S$ interchanged.

\section{Shape of spheroid with same surface area}
\label{app:spheroid}

For the spheroids described by Eq.~(\ref{eq:spheroid}) with surface area $S=4\pi R^2$
and a given eccentricity $e$, 
the semi-axes $A$ and $C$ are given as follows. 

For oblate spheroids with eccentricity $e$, the semi-axes $A$ and $C$ are
\begin{equation}
A
= R\sqrt{\frac{2}{1+\dfrac{1-e^2}{e}\operatorname{arctanh} e}}, 
\qquad
C = A\sqrt{1-e^2},
\end{equation}
whereas for prolate spheroids with eccentricity $e$, the semi-axes $A$ and $C$ are
\begin{equation}
A
= R\sqrt{\frac{2}{1+\dfrac{\arcsin e}{e\sqrt{1-e^2}}}},
\qquad
C = \frac{A}{\sqrt{1-e^2}}.
\end{equation}

\section{Approximate expressions of geodesic distance and parallel transport}
\label{app:approximation}

Here, approximate expressions for the geodesic distance between two points $\mathbf{x}_i$ and $\mathbf{x}_j$ and the parallel transport of a vector $\mathbf{v}_j$ at $\mathbf{x}_j$ to $\mathbf{x}_i$ along the geodesic between them are derived as follows.
Let $x^\alpha(s)$ denote the geodesic between $x_j^\alpha$ and $x_i^\alpha$, with $s\in[0,1]$, $x^\alpha(0)=x_j^\alpha$, and $x^\alpha(1)=x_i^\alpha$. 
It satisfies
\begin{equation}
    \frac{d^2x^\alpha}{ds^2}
    +\Gamma^\alpha_{\beta\gamma}\bigl(x(s)\bigr)\frac{dx^\beta}{ds}\frac{dx^\gamma}{ds}=0.
\end{equation}
With $\Delta^\alpha = x_j^\alpha - x_i^\alpha$ and 
the midpoint $\mathbf{m}=(\mathbf{x}_i+\mathbf{x}_j)/2$, 
the geodesic $x^\alpha(s)$ can be approximated as
\begin{equation}
x^\alpha(s)
=
x_j^\alpha
- s\,\Delta^\alpha
+\frac{1}{2}\Gamma^\alpha_{\beta\gamma}(\mathbf{m})\,s(1-s)\,\Delta^\beta\Delta^\gamma
+O(\lVert\Delta\rVert^3).
\label{eq:geodesicApproximation}
\end{equation}

The metric $g$ can be expanded around the midpoint using Eq.(\ref{eq:geodesicApproximation}) as
\begin{equation}
    g_{\alpha\beta}\bigl(x(s)\bigr)
    =
    g_{\alpha\beta}(\mathbf{m})
    +\left(\frac{1}{2}-s\right)\Delta^\ell\frac{\partial g_{\alpha\beta}(\mathbf{m})}{\partial x^\ell} 
    +O(\lVert\Delta\rVert^2).
\end{equation}
Combining this expansion with Eqs.~(\ref{eq:geodesicApproximation}) and (\ref{eq:metricDerivative}), one finds
\begin{equation}
g_{\alpha\beta}\bigl(x(s)\bigr)\frac{dx^\alpha(s)}{ds}\frac{dx^\beta(s)}{ds} 
= g_{\alpha\beta}(\mathbf{m})\,\Delta^\alpha\Delta^\beta 
+O(\lVert\Delta\rVert^4).
\end{equation}
Here, the third-order terms in $\Delta$ cancel. 
The approximated geodesic distance is then given by  
\begin{equation}
    \begin{aligned}
    d(\mathbf{x}_i,\mathbf{x}_j)
    &=
   \int_0^1 ds\,
   \sqrt{g_{\alpha\beta}\bigl(x(s)\bigr)\frac{dx^\alpha(s)}{ds}\frac{dx^\beta(s)}{ds}}
   \\
&= \sqrt{g_{\alpha\beta}(\mathbf{m})\,\Delta^\alpha\Delta^\beta}
   +O(\lVert\Delta\rVert^3).
    \end{aligned}
\end{equation}
The squared distance is obtained as Eq.~(\ref{eq:approximateDistance}).

Parallel transport $\tilde{\mathbf v}$ of $\mathbf{v}_j$ along the geodesic $x^\alpha(s)$ is obtained from
\begin{equation}
    \frac{d \tilde v^\alpha}{ds}
    +\Gamma^\alpha_{\beta \gamma}\bigl(x(s)\bigr)\frac{d x^\gamma}{ds}\tilde v^\beta=0,
    \label{eq:parallelTransport}
\end{equation}
with $\tilde v^\alpha(0)=v_j^\alpha$ and $\tilde v^\alpha(1)=\tilde v_{j,i}^\alpha$.
Using Eq.~(\ref{eq:geodesicApproximation}) together with
\begin{equation}
    \Gamma^\alpha_{\beta \gamma}\bigl(x(s)\bigr)
    =
    \Gamma^\alpha_{\beta \gamma}(\mathbf{m})
    -\left(s-\frac{1}{2}\right)
    \frac{\partial\Gamma^\alpha_{\beta \gamma}}{\partial x^\delta}(\mathbf{m})\,\Delta^\delta
    +O(\lVert\Delta\rVert^2),
\end{equation}
Eq.~(\ref{eq:parallelTransport}) can be solved to yield Eq.~(\ref{eq:parallelTransportApproximation}).

\bibliographystyle{apsrev4-2}

\bibliography{swarmOnCurvedSurface}

\clearpage
\onecolumngrid

\begin{itemize}
  \item \textbf{Movie 1:} Time evolution for Fig.~3(a) ($K=0.05$, $\sigma=0.2$).
  \item \textbf{Movie 2:} Time evolution for Fig.~3(b) ($K=0.1$, $\sigma=0.2$).
  \item \textbf{Movie 3:} Time evolution for Fig.~3(c) ($K=1.0$, $\sigma=0.2$).
  \item \textbf{Movie 4:} Time evolution for Fig.~4(a) (oblate spheroid; $e=0.5$, $K=0.2$, $\sigma=0.2$).
  \item \textbf{Movie 5:} Time evolution for Fig.~4(b) (oblate spheroid; $e=0.9$, $K=0.2$, $\sigma=0.2$).
  \item \textbf{Movie 6:} Time evolution for Fig.~4(c) (prolate spheroid; $e=0.5$, $K=0.2$, $\sigma=0.2$).
  \item \textbf{Movie 7:} Time evolution for Fig.~4(d) (prolate spheroid; $e=0.9$, $K=0.2$, $\sigma=0.2$).
\end{itemize}

\end{document}